\documentclass[conference]{IEEEtran}
\IEEEoverridecommandlockouts
\usepackage{cite}
\usepackage{amsmath,amssymb,amsfonts}
\usepackage{algorithmic}
\usepackage{graphicx}
\usepackage{subcaption}  
\usepackage{float} 
\usepackage{textcomp}
\usepackage{xcolor}
\usepackage{url}
\usepackage{placeins}
\def\BibTeX{{\rm B\kern-.05em{\sc i\kern-.025em b}\kern-.08em
    T\kern-.1667em\lower.7ex\hbox{E}\kern-.125emX}}
\begin{document}

\title{Adaptive Signal Analysis for Automated Subsurface Defect Detection Using Impact Echo in Concrete Slabs\\
\thanks{}
}

\author{
\IEEEauthorblockN{ Deepthi Pavurala}
\IEEEauthorblockA{\textit{School of Computing} \\
\textit{George Mason University}\\
Fairfax, VA, USA \\
dpavural@gmu.edu}
\and
\IEEEauthorblockN{Duoduo Liao}
\IEEEauthorblockA{\textit{School of Computing} \\
\textit{George Mason University}\\
Fairfax, VA, USA \\
dliao2@gmu.edu}
\and
\IEEEauthorblockN{ Chaithra Reddy Pasunuru}
\IEEEauthorblockA{\textit{School of Computing} \\
\textit{George Mason University}\\
Fairfax, VA, USA \\
cpasunur@gmu.edu}

}

\maketitle

\begin{abstract}
This pilot study presents a novel, automated, and scalable methodology for detecting and evaluating subsurface defect-prone regions in concrete slabs using Impact Echo (IE) signal analysis. The approach integrates advanced signal processing, clustering, and visual analytics to identify subsurface anomalies. A unique adaptive thresholding method tailors frequency-based defect identification to the distinct material properties of each slab. The methodology generates frequency maps, binary masks, and k-means cluster maps to automatically classify defect and non-defect regions. Key visualizations, including 3D surface plots, cluster maps, and contour plots, are employed to analyze spatial frequency distributions and highlight structural anomalies. The study utilizes a labeled dataset of eight reinforced concrete specimens constructed at the Federal Highway Administration (FHWA) Advanced Sensing Technology Nondestructive Evaluation Laboratory, each containing known artificial defects. Evaluations involve ground-truth masking, comparing the generated defect maps with top-view binary masks derived from the information provided by the FHWA. The performance metrics, specifically F1-scores and AUC-ROC, achieve values of up to 0.95 and 0.83, respectively. The results demonstrate the robustness of the methodology, consistently identifying defect-prone areas with minimal false positives and few missed defects. Adaptive frequency thresholding ensures flexibility in addressing variations across slabs, providing a scalable framework for detecting structural anomalies. Additionally, the methodology is adaptable to other frequency-based signals due to its generalizable thresholding mechanism. This automated and scalable pipeline minimizes manual intervention, ensuring accurate and efficient defect detection in structural health monitoring. Moreover, this approach holds potential for integrating multimodal sensor fusion in infrastructure maintenance and monitoring, further advancing Non-Destructive Evaluation (NDE) techniques.

\end{abstract}


\section{Introduction}
The structural health of concrete infrastructure, such as bridge decks and other critical public assets, is a growing concern due to the presence of subsurface defects that compromise their safety and durability\cite{gucunski2015, FHWA2021}. Non-Destructive Evaluation (NDE) methods have become indispensable for assessing concrete structures, enabling the detection of internal anomalies without causing damage\cite{FHWA2021}. Among these techniques, the Impact Echo (IE) method has shown promise in detecting subsurface flaws by analyzing frequency responses generated through mechanical impacts\cite{Sansalone_book1997, Sansalone1997,xu2017}. This capability makes IE highly suitable for inspecting concrete structures, particularly bridge decks, where defects often remain concealed.
However, the application of IE in practice faces significant challenges. Traditional approaches are often limited to single-layer conditions and rely on static, predefined threshold values for defect detection\cite{Sansalone_book1997, Azari2019}. These methods struggle to adapt to the unique characteristics of individual slabs, particularly in complex structures where material properties and environmental factors vary\cite{Qu2016}. Furthermore, existing studies on IE typically rely on manually defined thresholds that fail to account for variations in material and structural properties across different slabs, potentially leading to inaccurate detection and limited reproducibility.

This pilot study addresses these challenges by introducing a fully automated, frequency-adaptive methodology tailored to concrete structures. Our approach leverages adaptive frequency thresholds based on the unique frequency signatures of each slab, enabling precise detection of defect-prone areas. Instead of focusing on specific defect types, the methodology emphasizes identifying and localizing regions that may contain structural anomalies. The spatial mapping of these regions is enhanced through advanced visualizations, including contour plots, three-dimensional (3D) surface plots, cluster maps, and histogram plots, which provide interpretable insights into defect-prone areas.

This study uses the dataset sourced from the Federal Highway Administration (FHWA) Advanced Sensing Technology Nondestructive Evaluation laboratory, which comprises standardized, laboratory-controlled data on concrete slabs with known artificial defects designed to simulate bridge defects and facilitate the evaluation of structural anomalies using different NDE methods\cite{FHWA2021}. This dataset is characterized by its spatial structure and controlled conditions, which enables reliable analysis and validation of our adaptive methodology. By leveraging such a robust dataset, the study establishes a reliable foundation. This supports defect detection in concrete structures.

The key contributions of this pilot study are the development of a novel fully automated, frequency-adaptive methodology for detecting defect-prone areas in concrete slabs. By dynamically adjusting detection thresholds based on the frequency distributions observed in each slab, the approach enhances adaptability and accuracy without depending on material property inputs. Visualization techniques, such as contour maps and 3D surface plots, improve defect localization and interpretation. Furthermore, this study introduces a novel evaluation technique by creating a top-view representation of the concrete slab, derived from information provided by the FHWA, which serves as ground truth for evaluating defect detection results. Binary contour and cluster plots, generated through the automated methodology, are masked and aligned with this Ground Truth Mask (GTM) to calculate key evaluation metrics, including Intersection over Union (IoU), Precision, Recall, F1-score, False Positive Rate (FPR), False Negative Rate (FNR), and Area Under the Receiver Operating Characteristic Curve (AUC-ROC). This rigorous masking and evaluation process ensures the reliability of the methodology and provides quantitative validation of its performance. By integrating these techniques, the study establishes a scalable framework for broader NDE applications, including slabs with or without overlays and other frequency-based signals, leveraging adaptive frequency thresholding for robust defect detection.

\section{Related Work}

NDE techniques are essential for detecting defects in concrete structures, such as bridge decks, slabs, and walls. Various NDE methods, including IE, Ground-Penetrating Radar (GPR), Ultrasonic Surface Wave (USW), and Electrical Resistivity (ER), are widely used to assess the health of concrete structures, especially for detecting delamination, cracks, and corrosion of reinforcing steel \cite{gucunski2015, FHWA2021}. Among these, the IE method stands out for its ability to detect subsurface anomalies by measuring frequency shifts of transient stress waves \cite{Sansalone1997}. This method is particularly effective in detecting delamination and cracks, with several studies detailing its application to bridge decks and slabs \cite{Sansalone1997, gucunski2015}.

The use of IE signals for bridge defect detection has been extensively studied, employing various advanced techniques, such as signal transformation \cite{Qu2016, Kim2020}, Variational Mode Decomposition (VMD) \cite{xu2017}, frequency analysis \cite{liu2019, sajid2021, pandrum2024}, statistical pattern recognition \cite{sajid2021}, and Machine Learning (ML) \cite{pandrum2024,Azari2019, dorafshan2020,transfer2023,classification2023}.  The IE method is also widely applied in  assessing concrete slabs, as documented in FHWA studies \cite{sengupta2024, FHWA2021, gucunski2015}. These studies identify subsurface anomalies such as delamination, cracks, and voids through structured grid patterns and frequency response analysis, correlating frequency shifts with structural anomalies. Despite its strengths, IE faces challenges such as sensitivity to noise and reliance on predefined frequency thresholds, which our work addresses through dynamic frequency thresholds and clustering-based defect detection.

\subsection{Frequency Analysis for IE Signals}

Frequency analysis, particularly using Fast Fourier Transform (FFT), is extensively used to identify specific frequencies linked to defects such as cracks, voids, and delamination \cite{pandrum2024}. Studies, such as those by Liu et al. \cite{liu2019} and Sajid et al. \cite{sajid2021}, demonstrate the effectiveness of spectral response analysis for detecting and categorizing defects. Additionally, methods like the Extreme Studentized Deviate (ESD) test and Principal Component Analysis (PCA) are used to classify defect regions based on their frequency characteristics \cite{sajid2021}. 

In our study, we utilize FFT for frequency analysis and introduce dynamic thresholds tailored to each slab, enabling the precise identification of defect-prone areas within concrete slabs. This approach diverges from prior studies that focus on static frequency peak identification for general classifications by dynamically adjusting frequency ranges to accommodate slab-specific variations, thereby enhancing the detection of localized structural anomalies.

\subsection{ML for IE Signals}
Recent advancements in ML \cite{pandrum2024, dorafshan2020} enhance the application of IE in bridge defect detection based on labeled IE data.\cite{transfer2023} applies transfer learning to adapt laboratory-trained models for field data, addressing the scarcity of labeled field data. \cite{classification2023} explores explainable deep learning and transfer learning, emphasizing model interpretability in NDE applications.

Additionally, clustering techniques, as key unsupervised ML methods, play a significant role in defect classification for NDE applications. For instance, \cite{sengupta2024} uses unsupervised clustering to classify IE signals from real bridge deck data (the FHWA InfoBridge dataset) into categories (Good, Fair, and Poor), validating a physics-based scheme with frequency partitioning. \cite{gongzhang2016} proposes a clustering algorithm for ultrasonic NDE, enhancing defect detection by separating real defects from noise. Similarly, \cite{chen2021} combines K-means clustering with the Level Set Method for low-contrast segmentation, offering valuable insights for improving noisy data classification. While prior studies have successfully leveraged frequency analysis, clustering techniques, and ML for defect detection, our work introduces a novel combination of dynamic frequency thresholding and unsupervised clustering tailored to each slab’s unique frequency characteristics. This approach improves detection accuracy and avoids reliance on large labeled datasets, distinguishing it from both traditional and ML-based methods.

\quad
\quad

Building on this foundation, we leverage FFT-based frequency analysis and clustering techniques to identify defect-prone areas within concrete slabs. In contrast to prior studies emphasizing static frequency thresholds or generalized peak identification for defect classification, our method introduces dynamic frequency range adjustments tailored to the specific characteristics of each slab. This tailored approach enhances the accuracy of detecting potential structural anomalies. Additionally, we integrate advanced spatial visualizations, including 3D surface plots and contour maps, to provide intuitive, detailed representations of defect-prone regions. These visual tools not only aid in the effective mapping of anomalies but also improve the interpretability and usability of the analysis in practical applications.

\section{Methodology}
\begin{figure}[h]
    \centering
    \includegraphics[width=0.95\linewidth]{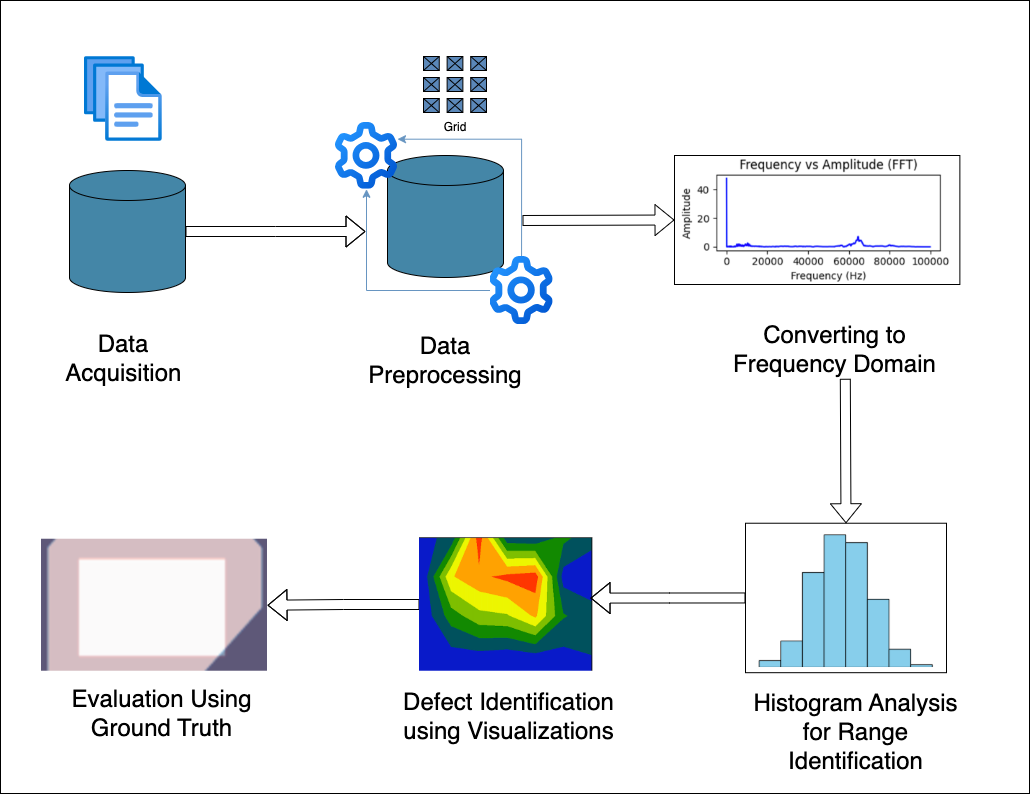}
    \caption{IE data analysis workflow for defect detection}
    \label{figure: Methodology}
\end{figure}

This study utilizes an automated process for defect detection in concrete slabs through the analysis of the IE signals. Figure \ref{figure: Methodology} outlines our methodology, illustrating the key stages of data acquisition, signal processing, adaptive thresholding, defect detection, and evaluation metrics. It highlights the integration of these steps to detect subsurface defects in concrete slabs, with a focus on the IE method.

\subsection{Data Acquisition}
The dataset for this analysis is sourced from the FHWA report titled \textit{Nondestructive Evaluation of Concrete Bridge Decks without Overlays} \cite{FHWA2021}. This report provides an extensive dataset collected using NDE methods to detect structural anomalies in concrete bridge decks without overlays. We focus on IE data, which effectively identifies subsurface defects by detecting changes in material density and structural integrity. The dataset consists of 252 individual IE readings per slab, recorded across a 9x28 spatial grid. These readings serve as the foundation for further frequency-based analysis, including the identification of peak frequencies for defect detection.
\subsection{Data Processing}
\subsubsection{Data Normalization}

The amplitude values in each signal are normalized to a standard scale. This normalization adjusts the amplitude values of each signal to lie between 0 and 1, using the formula:

\[\small A_{\text{normalized}} = \frac{A(t) - \min(A)}{\max(A) - \min(A)}
\]
where $A(t)$ represents amplitude, a time-domain signal.

This normalization reduces the influence of any external factors that could cause variations in amplitude during data collection, enabling a focus on the frequency characteristics of each signal for defect detection.

\subsubsection{Frequency Transformation}
Once normalized, each amplitude signal is analyzed in the frequency domain by applying the Fast Fourier Transform (FFT) \cite{Kumar2019}. FFT converts the time-domain signal into its frequency components, making it possible to detect dominant frequencies that correlate with specific defect types \cite{Azari2019}. For each amplitude signal \( A(t) \), FFT provides the frequency components:

\[\small F(f) = \text{FFT}(A(t))\]

Since the FFT output is symmetric, negative frequencies are redundant and only positive frequencies are retained, as they contain the relevant information for defect detection.

The frequency with the highest spectral amplitude within the identified frequency range is used as the \textit{dominant frequency} for each grid point, which is then classified into defect-prone or intact regions.

\subsection{Adaptive Frequency Thresholding}
The Adaptive Frequency Thresholding method analyzes histogram distributions to automatically identify clustered frequency ranges, determining defect-prone areas from intact regions by evaluating the distribution of dominant frequencies across all measurement points on a slab. This dynamic adjustment of thresholds improves defect detection accuracy by tailoring the analysis to the specific characteristics of the data. 
\subsubsection{Histogram Generation}

The bin size determines the resolution of frequency visualization, crucial for identifying defect-prone areas. In this study, the bin size $k$ is calculated as the mean of the \textit{Exponential Rule} and \textit{Adjusted Square Root Rule}\cite{DeLaRubia2024}, ensuring adaptability and statistical robustness.

\paragraph{Exponential Rule}
The Exponential Rule defines the number of bins $k$ as:
\[
k = \lceil n^{1 / \text{exponent}} \rceil,
\]
where $n$ is the total number of data points, and the exponent is typically set to 1.5.

\paragraph{Adjusted Square Root Rule}
This method modifies the classic square root rule:
\[
k = \lceil \text{multiplier} \cdot \sqrt{n} \rceil,
\]
where the multiplier (e.g., 1.5) adjusts the bin size based on dataset variability.

\paragraph{Combined Approach}
The bin size $k$ is calculated as:
\[
k = \text{mean}(\textit{Exponential Rule}, \textit{Adjusted Square Root Rule}).
\]

\paragraph{Histogram Construction}
Using $k$ bins, the histogram density $H(b)$ for bin $b$ is:
\[
H(b) = \frac{\text{Number of } f_i \text{ in bin } b}{N \cdot \Delta f},
\]
where $f_i$ represents the dominant frequency, $N$ is the total data points, and $\Delta f$ is the bin width:
\[
\Delta f = \frac{\text{Range of frequencies}}{k}.
\]

This combined method ensures effective frequency distribution representation while maintaining statistical consistency.

\subsubsection{Frequency Range Identification} 

Frequency ranges are identified by grouping adjacent bins in the histogram where \( H(b) > 0 \). Starting from the first bin with \( H(b) > 0 \), the range is expanded until a bin with \( H(b) = 0 \) is encountered, marking the end of the region. The frequency range \([f_{\text{start}}, f_{\text{end}}]\) is given by:

\[ \small
[f_{\text{start}}, f_{\text{end}}] = \big[f_{\text{min}} + (b_s \cdot \Delta f), \, f_{\text{min}} + ((b_e + 1) \cdot \Delta f) \big],
\]

where \( b_s \) and \( b_e \) are the indices of the first and last bins with \( H(b) > 0 \), and \( \Delta f = \frac{f_{\text{max}} - f_{\text{min}}}{k} \) is the bin width.

\subsubsection{Frequency Range Classification}
The identified frequency ranges are categorized into \textit{low-frequency ranges} (indicative of defects) and \textit{high-frequency ranges} (indicative of intact areas). \textit{Low-frequency range} is the minimum range $[f_{\text{low-start}}, f_{\text{low-end}}]$ chosen to represent defect-prone areas. \textit{High-frequency range} is the maximum range $[f_{\text{high-start}}, f_{\text{high-end}}]$ chosen to represent intact areas.

FFT outputs are filtered to analyze only frequencies within the identified ranges. The frequency values within these ranges are used to classify each grid point as either defect or intact.

\subsection{Position Mapping}
Position mapping is the process of associating frequency data with specific spatial locations on a concrete slab. The dataset consists of measurements taken across a 9x28 grid, which represents the spatial layout of the slab. Each grid cell corresponds to a particular measurement point on the slab's surface, ensuring that every frequency reading is tied to a specific spatial position.

Each measurement point's dominant frequency is mapped to its respective grid location. The frequencies are classified as either low (defect-prone) or high (intact) based on the results of adaptive frequency thresholding. This classification ensures that each grid cell contains either a low or high peak frequency, indicating whether the area is defect-prone or intact.

This mapping allows for a detailed spatial representation of the slab, where defects can be visually identified based on their frequency characteristics. By associating frequency data with specific grid positions, the mapping process provides an effective means of visualizing the distribution of defects across the slab’s surface, aiding in the detection and analysis of structural anomalies.

\subsection{Defect Identification}

\subsubsection{Defect Identification through Contour Mapping} Contour maps are generated from the \textit{low--frequency range} to analyze and visualize frequency characteristics and defect regions in concrete slabs. It provides a spatial representation of frequency characteristics across the slab surface. Areas with anomalies typically show as deviations from the expected frequency range and are highlighted using color.

Binary contour mapping processes the classification of each grid point as either defect (\(1\)) or non-defect (\(0\)) based on dominant frequency ranges identified through adaptive thresholding. Several techniques can be employed to automatically determine the binary threshold. For instance, in this pilot study, binary classification is performed by setting the threshold based on the median of low frequencies for each slab. A grid point is classified as a defect (1) if its frequency is below the threshold, and as non-defect (0) otherwise. Let \( f_{\text{threshold}} \) be the threshold frequency. The binary classification \( B(x, y) \) at position \((x, y)\) is given by:
\[\small
B(x, y) =
\begin{cases} 
1, & f(x, y) < f_{\text{threshold}} \\
0, & f(x, y) \geq f_{\text{threshold}}
\end{cases}
\]

 These values are mapped onto the spatial grid, highlighting defect and non-defect regions.

\subsubsection{Defect Identification through Clustering}

Cluster maps are generated using K-means clustering to group grid points into defect and non-defect clusters based on dominant frequencies in low-frequency range, as determined by adaptive thresholding. The clusters are formed by minimizing the within-cluster variance:
\[\small
\text{Cost} = \sum_{k=1}^{K} \sum_{x_i \in C_k} \| x_i - \mu_k \|^2
\]
where \( K \) is the number of clusters (e.g., \( K = 2 \) for defect vs. non-defect), \( C_k \) represents the cluster \( k \), 
    \( \mu_k \) represents the mean of cluster \( k \), and \( x_i \) represents a data point (frequency) in cluster \( k \).

The \textit{cluster centroids} represent the mean frequency value for each cluster. By comparing these centroids, the defect cluster is identified by the lowest centroid frequency, which typically corresponds to the low-frequency region indicating defects, while the higher centroid corresponds to the intact region.

\subsection{Evaluation}

\begin{figure}[h]
    \centering
    \includegraphics[width=0.9\linewidth]{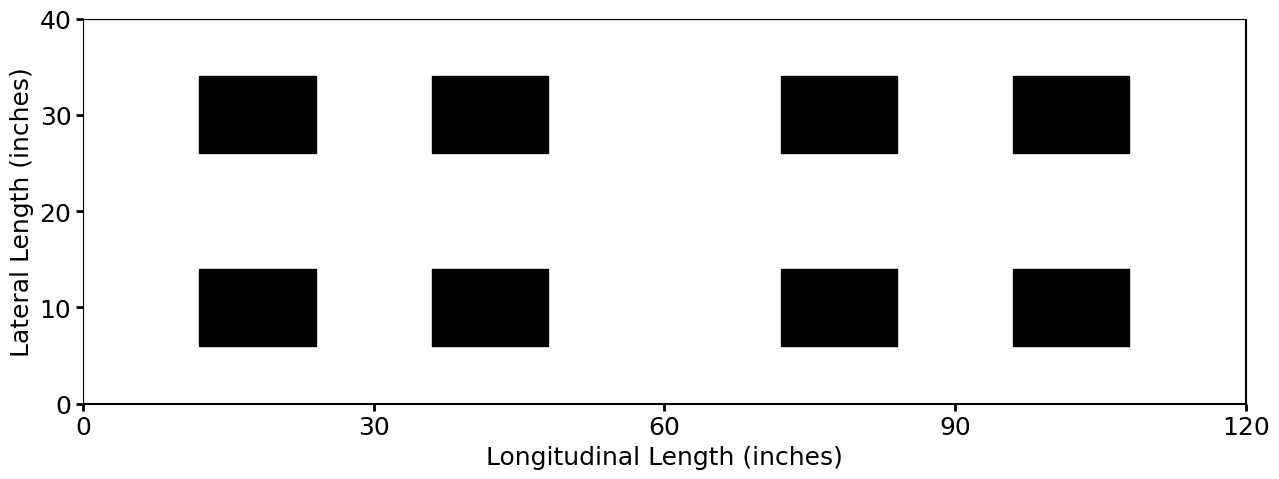}
    \caption{Slab top view with defect representation as per FHWA}
    \label{figure:slab top view}
\end{figure}

The evaluation methodology for detecting defects in concrete slabs involves creating a top-view representation of the slab with defects based on the lab specimen report published by FHWA \cite{FHWA2021_web} and comparing it against binary plots generated during analysis. The comparison is based on performance metrics, including IoU, Precision, Recall, F1 score, FNR, FPR, TNR, and AUC-ROC. These metrics help assess the accuracy of defect identification in the binary plots.
Below is the representation of defects added to slabs in Figure~\ref{figure:slab top view}. The defects are located beneath the surface within the slab.

\subsubsection{Creating A GTM for the Concrete Slab}

Figure \ref{figure:slab top view} illustrates a generated top-down view of the slab,  used to visualize the known defect positions. Defects are represented as rectangles on a grid, positioned based on the slab dimensions and defect specifications. This top-view image serves as a ground truth mask for evaluating the binary defect plots.

\subsubsection{Evaluation of Binary Defect Plots}

The binary plots from defect detection, referred as Detect Mask (DM), are compared to the GTM using pixel-wise comparisons, where 1 represents no defect (white pixel) and 0 represents a defect (black pixel). True Positives (TP) are pixels where both DM and GTM are 0. False Positives (FP) occur when a pixel is 0 in DM but 1 in GTM, and False Negatives (FN) occur when a pixel is 1 in DM but 0 in GTM. Edge cases, such as slabs with no defect pixels in the GTM, are handled to prevent division by zero and ensure accurate metrics.

The following metrics \cite{rainio2024, rezaie2024} are employed for evaluation, specifically tailored to assess defect detection in this study:

\paragraph{Intersection over Union (IoU)}
IoU measures the overlap between the DM and the GTM. It is calculated as:
\[
\small
\text{IoU} = \frac{\sum (\text{GTM} \cap \text{DM})}{\sum (\text{GTM} \cup \text{DM})}
\]
Here, \(\cap\) denotes the intersection, and \(\cup\) denotes the union of the masks.

\paragraph{Precision}
Precision quantifies the proportion of correctly identified defect pixels out of all detected defect pixels:

\[
\small
\text{Precision} = \frac{\text{TP}}{\text{TP}+\text{FP}}
\]

\paragraph{Recall}
Recall evaluates the proportion of correctly identified defect pixels out of all ground truth defect pixels:

\[
\small
\text{Recall} = \frac{\text{TP}}{\text{TP}+\text{FN}}
\]

\paragraph{F1 Score}
The F1 score provides a harmonic mean of precision and recall:
\[
\small
\text{F1} = 2 \cdot \frac{\text{Precision} \cdot \text{Recall}}{\text{Precision} + \text{Recall}}
\]

\paragraph{False Negative Rate (FNR)}
The FNR represents the proportion of actual defect pixels that the model fails to detect i.e., the proportion of defect pixels in the GTM that are missing from DM:
\[
\small
\text{FNR} = \frac{\text{FN}}{\text{FN}+\text{TP}}
\]  

\paragraph{False Positive Rate (FPR)}
The FPR represents the proportion of non-defect pixels incorrectly classified as defects i.e., the proportion of defect pixels in the DM that are missing from GTM:
\[
\small
\text{FPR} = \frac{\text{FP} }{\text{FP}+\text{TN}}
\]  

\paragraph{True Negative Rate (TNR)}
The TNR, also known as specificity, measures the proportion of non-defect pixels that were correctly identified as non-defects i.e., the number of non-defect pixels identified in both GTM and DM:
\[
\small
\text{TNR} = \frac{\text{TN}}{\text{TN}+\text{FP}}
\]  

\paragraph{AUC-ROC}
The AUC-ROC measures the ability of the model to distinguish between defects and non-defects across various thresholds. It is calculated as the area under the curve formed by plotting the True Positive Rate (TPR) or Recall against the FPR for different decision thresholds. An AUC-ROC value of 1 indicates perfect classification, while a value of 0.5 suggests performance equivalent to randomization.

\normalsize

\begin{figure}[b]
    \centering
    \includegraphics[width=0.9\linewidth]{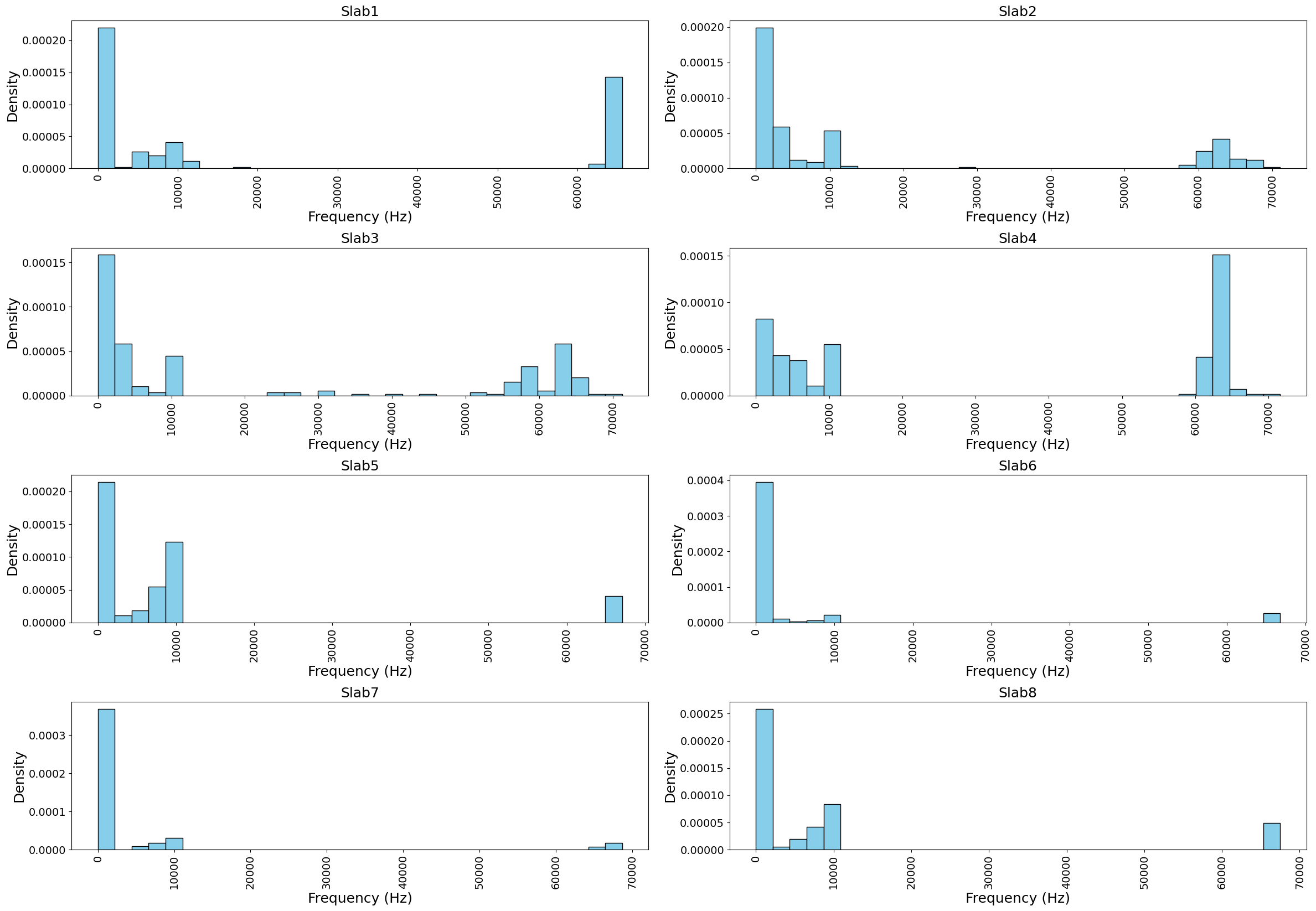}
    \caption{Frequency distribution for all slabs}
    \label{figure:Histograms}
\end{figure}

\begin{figure*}[h]
    \centering
    \includegraphics[width=0.8\linewidth]{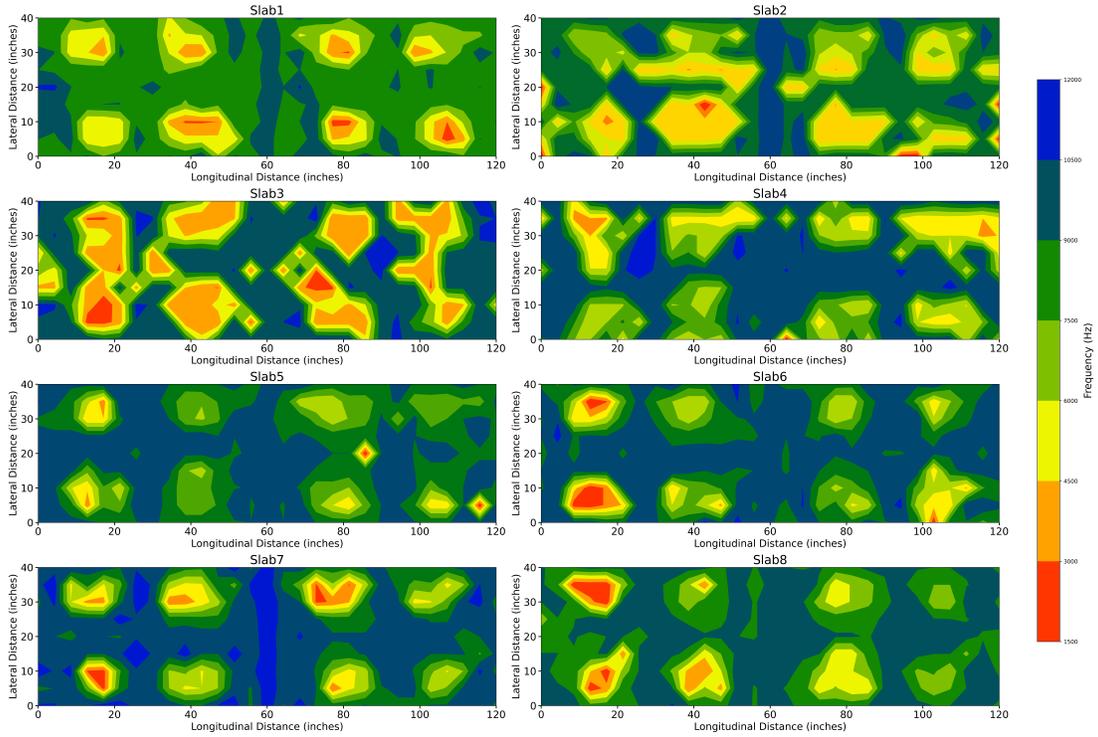}
    \caption{Contour plots for all slabs}
    \label{figure:contour all_slabs}
\end{figure*}

\section{Experimental Results and Data Analysis}
In this section, we present the experimental results from the analysis of the concrete slabs using IE signals to identify defects. Frequency values are mapped across grids and visualized through histograms, contour maps, cluster maps, and 3D surface plots. The experiments utilize Google Colab and Python libraries such as NumPy, SciPy, Matplotlib, and Scikit-learn for signal processing and visualization.

\begin{table}[h]
    \centering
    \begin{tabular}{|c|c|c|}
        \hline
        \textbf{Slab} & \textbf{Low Frequency Range (Hz)} & \textbf{High Frequency Range (Hz)} \\
        \hline
        Slab1 & (0.0, 12709.48) & (61429.17, 65665.67) \\
        Slab2 & (0.0, 13775.07) & (57396.11, 71171.17) \\
        Slab3 & (0.0, 11511.51) & (50650.65, 71371.37) \\
        Slab4 & (0.0, 11559.95) & (57799.74, 71671.67) \\
        Slab5 & (0.0, 10833.41) & (65000.48, 67167.17) \\
        Slab6 & (0.0, 10784.98) & (64709.87, 66866.87) \\
        Slab7 & (4436.69, 11091.74) & (64332.07, 68768.77) \\
        Slab8 & (0.0, 10897.99) & (65387.97, 67567.57) \\
        \hline
    \end{tabular}
    \caption{Frequency ranges identified automatically for each slab}
    \label{table:updated_frequency_ranges}
\end{table}

\subsection{Frequency Distribution and Analysis}
The histograms depict the frequency distribution across all measurements for each slab, with notable peaks at certain frequency intervals. The presence of high counts at lower frequencies generally corresponds to defect regions, as defects typically alter the frequency response of the material. The presence of multiple peaks or high counts at higher frequencies may indicate areas that are defect-free. By comparing the frequency distribution, it can be inferred the relative severity and types of defects present in different slabs. For instance, slabs with dominant lower-frequency peaks may indicate significant defect presence, while more uniform distributions suggest a more consistent, possibly defect-free structure.
Figure~\ref{figure:Histograms} shows the histograms for frequency range identification, with bin sizes determined as the mean of the \textit{Exponential Rule} and \textit{Adjusted Square Root Rule}. This approach ensures robust resolution and highlights two prominent frequency clusters in the lower and higher ranges for defect detection.

\subsection{Identified Frequency Ranges}
The frequency ranges for each slab are identified automatically using proposed adaptive thresholding techniques, as presented in Table~\ref{table:updated_frequency_ranges}. These ranges are used for defect detection. The low-frequency ranges indicate defect-prone areas, while the high-frequency ranges correspond to intact regions.

\subsection{Defect Region Detection}
\subsubsection{Contour Mapping for Defect Identification}
 Figure~\ref{figure:contour all_slabs} displays contour maps for eight slabs, with each plot indicating frequency distributions across the longitudinal (x-axis) and lateral (y-axis) dimensions of the slab. The color bar on the right shows the frequency in Hertz (Hz), ranging from approximately 1500 Hz to 12000 Hz. Lower frequencies are represented by warmer colors (red, orange, and yellow) and are generally indicative of potential defect-prone areas, while higher frequencies, represented by cooler colors (green and blue), suggest regions with intact material.

In these plots, Slabs 2, 3, and 4 stand out with significant areas of red and orange, particularly concentrated in the left and central portions, which may correspond to larger or more numerous defects. These slabs show a prominent contrast between defect-prone areas and intact regions, indicating higher structural variability. In contrast, Slabs 1, 5, 6, 7, and 8 display fewer low-frequency zones, with the green and blue colors covering more of their surfaces. This indicates fewer defects and a more uniform material condition across these slabs. Overall, these contour maps provide a valuable visual assessment of structural integrity, highlighting areas of potential defects in each slab. These frequency-based visualizations provide a clear spatial alignment with ground truth data from the FHWA report \cite{FHWA2021_web}, successfully identifying most defect positions. However, certain defects reported by the FHWA remain undetected, highlighting potential limitations in detection sensitivity or threshold settings. Refining these parameters could enhance the methodology’s ability to capture more subtle defect types, ensuring a more comprehensive match with the ground truth \cite{FHWA2021_web}.

\begin{figure}[t]
    \centering
    \includegraphics[width=0.98\linewidth]{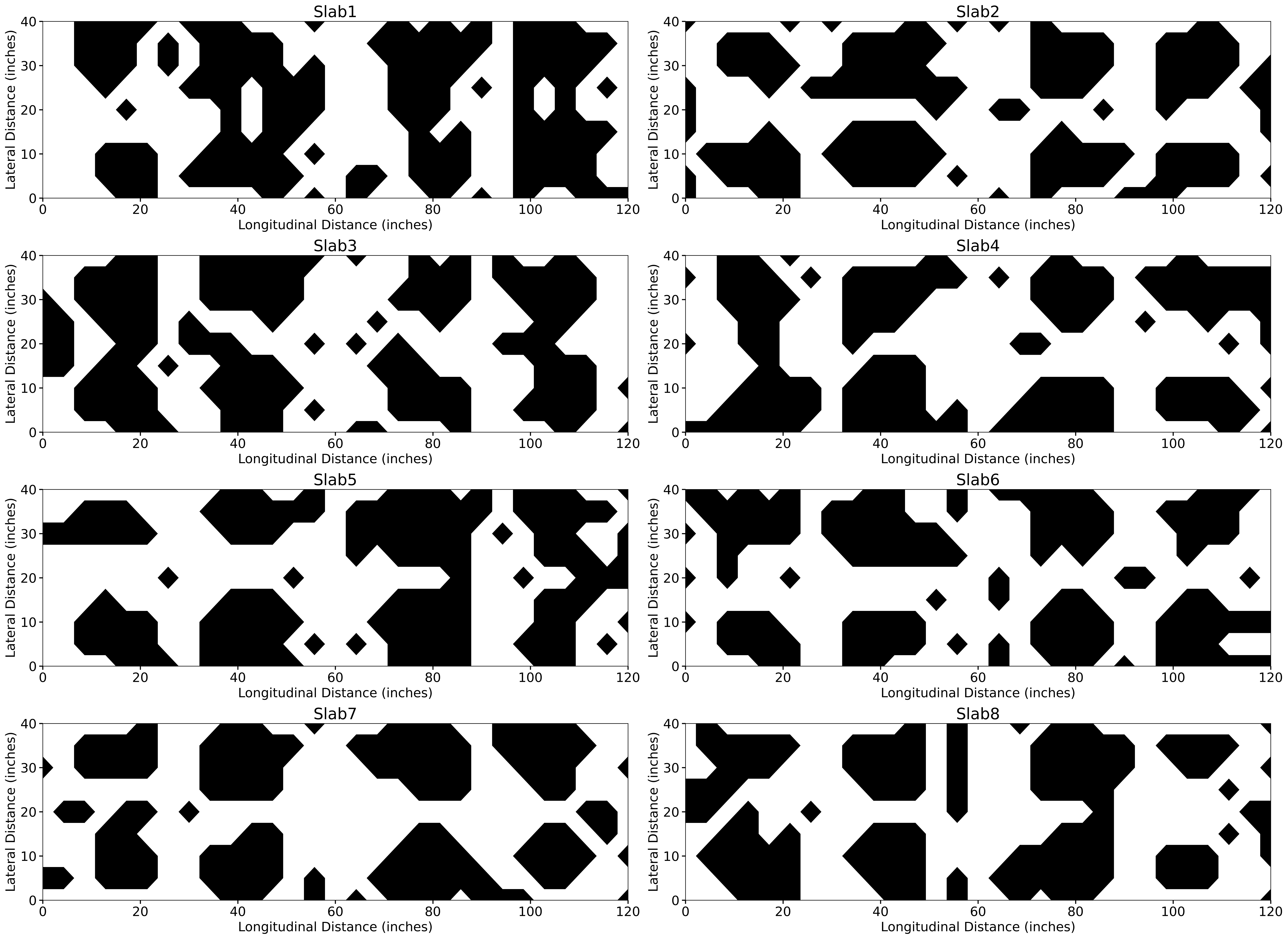}
    \caption{Binary masks for all slabs}
    \label{figure:binary contour all_slabs}
\end{figure}

\subsubsection{Binary Contour Mapping for Defect Identification}

The binary contour maps in Figure~\ref{figure:binary contour all_slabs} provide a clear visualization of potential defect areas across eight slabs, with black regions representing defect-prone zones and white regions indicating intact material. The distribution of black areas varies across the slabs, with Slabs 1, 3, 4, 5, and 8 showing more extensive defect regions, particularly in central and lower lateral sections, suggesting a higher concentration of structural anomalies in these slabs. In contrast, Slabs 2, 6, and 7 display fewer and more scattered black regions, indicating relatively fewer defects and a larger proportion of intact material.

This binary representation aids in quickly identifying defect patterns and prioritizing areas for further inspection or maintenance. The concentration of defects in certain slabs, as indicated by larger black regions, suggests these areas may require more immediate attention compared to slabs with predominantly white regions.

\subsubsection{Cluster Maps for Defect Identification}

Figure~\ref{figure:Cluster Maps all_slabs} illustrates the cluster maps for each slab (Slab1 through Slab8), generated using K-means clustering with two clusters. These maps show distinct frequency regions associated with material conditions. Darker areas indicate low-frequency clusters, which are often linked to defect-prone zones, while lighter areas represent higher-frequency clusters, suggesting intact or sound regions. The distribution of defects varies across slabs, with Slabs 2, 3, and 4 displaying more extensive defect regions, particularly concentrated in central and lateral sections. Conversely, Slabs 1, 5, 6, 7, and 8 exhibit smaller, more isolated clusters of defects, suggesting fewer structural concerns in these slabs. This aligns with exact position of defects with ground truth. This clustering approach highlights spatial variations in frequency response across each slab, supporting targeted identification of defect-prone regions based on frequency analysis.

\subsubsection{3D Surface Visualization}

\begin{figure*}[h]
    \centering
    \includegraphics[width=0.8\linewidth]{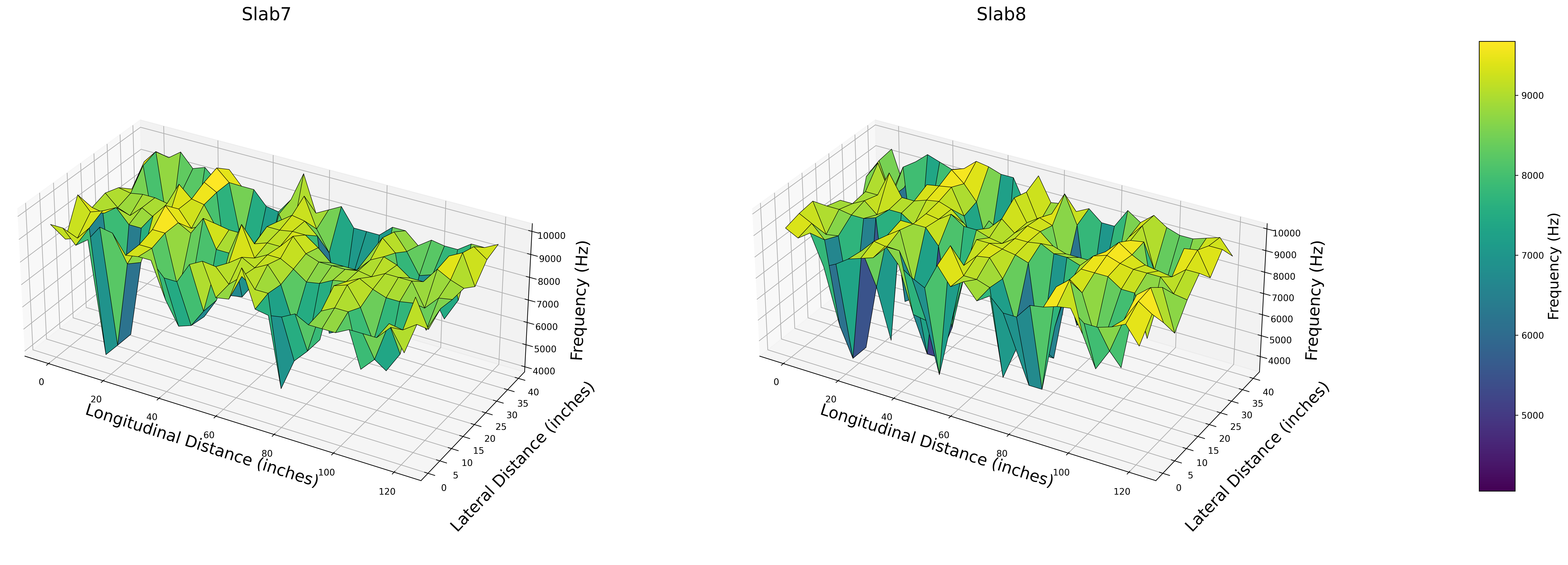}
    \caption{3D surface plots for Slab 7 and Slab 8}
    \label{figure:3D_surface_plots_slab_7_8}
\end{figure*}

Figure~\ref{figure:3D_surface_plots_slab_7_8} illustrates the 3D surface visualizations for slabs 7 and 8, mapping the distribution of frequency peaks across the slab surfaces.High-frequency peaks (yellow regions) represent intact material, while low-frequency valleys (blue and green) indicate potential defects.Slab 7 shows relatively smooth frequency distributions, particularly in the central sections, suggesting fewer structural concerns. In contrast,  Slab 8 exhibits more noticeable valleys across all sections, pointing to potential irregularities.

\begin{figure}[t]
    \centering
    \includegraphics[width=1\linewidth]{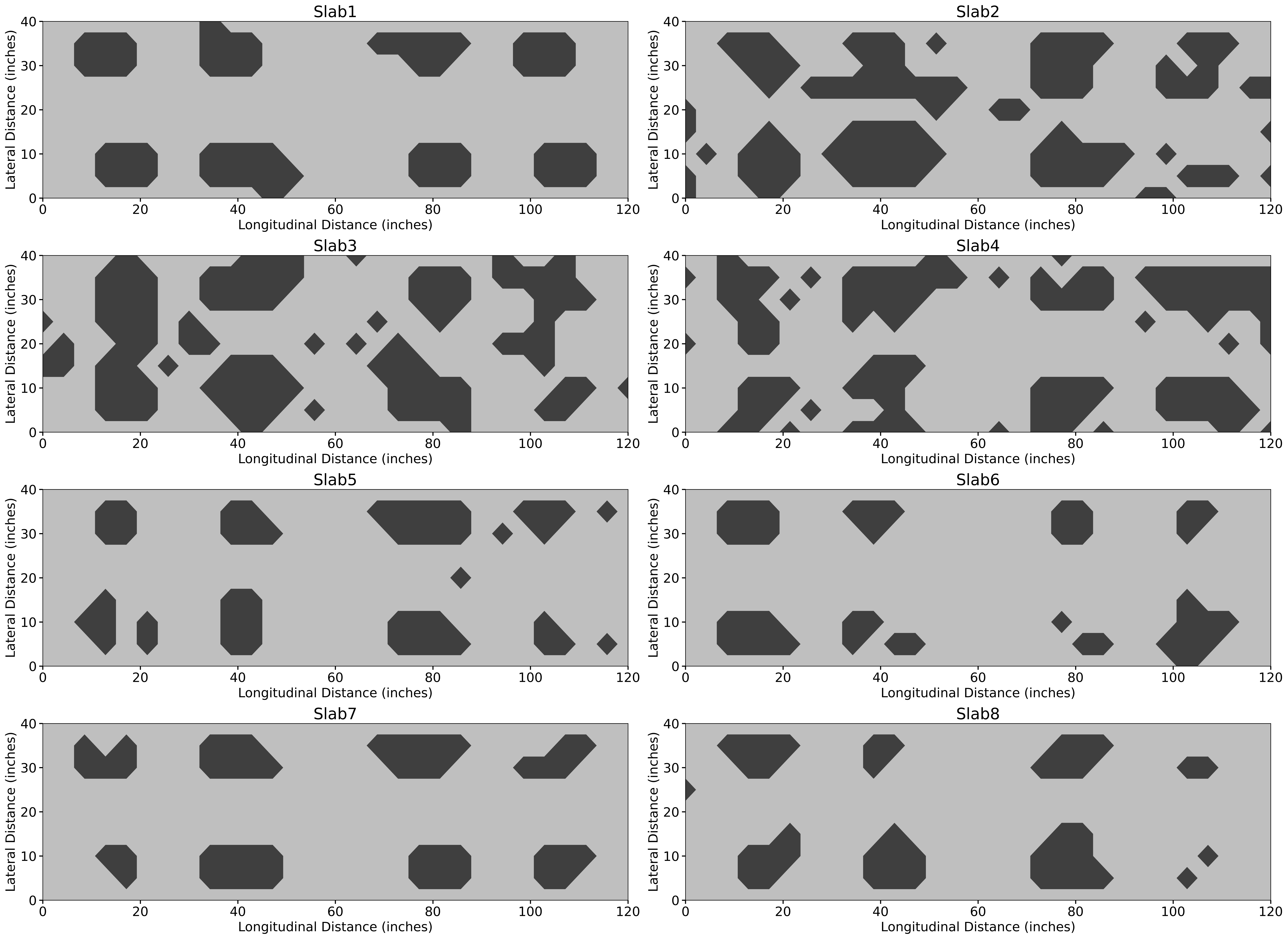}
    \caption{Cluster maps for all slabs}
    \label{figure:Cluster Maps all_slabs}
\end{figure}

In comparison, the binary contour maps in Figure~\ref{figure:binary contour all_slabs} highlight defect-prone areas as black regions, with Slab 7 showing scattered defects and Slab 8 displaying larger, concentrated regions of concern. Similarly, the cluster maps in Figure~\ref{figure:Cluster Maps all_slabs} confirm this pattern, with Slab 8 exhibiting more extensive low-frequency clusters compared to the smaller, isolated clusters in Slab 7. These visualizations collectively provide a detailed and corroborated view of defect distribution, enabling targeted inspection and maintenance.

The 3D surface visualizations provide detailed spatial insights into frequency variations, supporting the targeted identification and localization of potential defects within each slab.

\subsection{Experimental Summary}

The visualizations provide crucial insights into the distribution of defects across slabs based on frequency analysis. Contour maps highlight prominent low-frequency zones (red/yellow shades), identifying defect-prone areas. Slabs 2 and 3 exhibit extensive low-frequency regions in their central and lateral sections, indicating a higher concentration of structural anomalies. In contrast, Slabs 1, 4, 5, 6, 7, and 8 show fewer low-frequency areas, suggesting better structural integrity. Binary contour maps reinforce these findings, with dark regions clearly delineating defect-prone zones and light regions representing intact material. Slabs 2, 3, and 4 display more extensive dark areas, reflecting a higher defect density, whereas Slabs 1, 5, 6, 7, and 8 exhibit smaller and more scattered dark regions, indicative of relatively fewer defects.

Cluster maps, generated using K-means clustering, provide additional insights into defect distribution. Darker areas represent low-frequency clusters linked to defects, while lighter areas signify intact material. Slabs 2, 3, and 4 reveal larger and more concentrated defect clusters, particularly in their central and lateral regions, aligning with observations from contour and binary maps. Slabs 1, 5, 6, 7, and 8 exhibit smaller and more isolated clusters, reflecting a lower density of structural anomalies. This clustering approach effectively emphasizes spatial variations in defect-prone regions, complementing the binary and contour map analyses.

Overall, the study identifies Slabs 2, 3, and 4 as the most defect-prone, requiring immediate attention due to their extensive defect zones and high concentration of structural anomalies. Slabs 1, 5, 6, 7, and 8 are relatively intact, with fewer and more scattered defect zones, indicating better structural soundness. These visualizations and analyses collectively enable targeted prioritization of interventions based on defect severity and distribution.

\section{Evaluation and Error Analysis}

\subsection{Evaluation Results}

Table~\ref{table:Performance Metrics for Contour Slabs} summarizes the performance metrics for each slab using binary contour plot, while  Table~\ref{table:Performance Metrics for Cluster Maps} presents the corresponding metrics for Cluster maps. These values are computed by overlaying the detection results onto the ground truth mask in Figure \ref{figure:slab top view} for each slab, enabling pixel-by-pixel analysis.
Figures~\ref{figure:binary contour all_slabs} and \ref{figure:Cluster Maps all_slabs} illustrate the binary contour plots and cluster maps used in the evaluation.

\begin{table}[b]
    \centering
    \renewcommand{\arraystretch}{1.2} 
    \setlength{\tabcolsep}{5pt} 
    \scriptsize 
    \begin{tabular}{|c|c|c|c|c|c|c|c|c|}
        \hline
        \textbf{Slab} & \textbf{IoU} & \textbf{Precision} & \textbf{Recall} & \textbf{F1} & \textbf{FNR} & \textbf{FPR} & \textbf{TNR} & \textbf{ROC} \\
                \hline
        Slab1 & 0.74 & 0.97 & 0.75 & 0.85 & 0.25 & 0.17 & 0.83 & 0.79 \\
        Slab2 & 0.78 & 0.98 & 0.79 & 0.88 & 0.21 & 0.13 & 0.87 & 0.83 \\
        Slab3 & 0.75 & 0.98 & 0.76 & 0.86 & 0.24 & 0.14 & 0.86 & 0.81 \\
        Slab4 & 0.77 & 0.98 & 0.78 & 0.87 & 0.22 & 0.15 & 0.85 & 0.82 \\
        Slab5 & 0.75 & 0.97 & 0.77 & 0.86 & 0.23 & 0.19 & 0.80 & 0.78 \\
        Slab6 & 0.77 & 0.98 & 0.79 & 0.87 & 0.21 & 0.20 & 0.80 & 0.79 \\
        Slab7 & 0.78 & 0.98 & 0.80 & 0.88 & 0.20 & 0.13 & 0.87 & 0.83 \\
        Slab8 & 0.77 & 0.97 & 0.79 & 0.87 & 0.21 & 0.19 & 0.81 & 0.80 \\
        \hline
        
    \end{tabular}
    \caption{Performance metrics for binary masks}
    \label{table:Performance Metrics for Contour Slabs}
\end{table}

\begin{table}[b]
    \centering
    \renewcommand{\arraystretch}{1.2} 
    \setlength{\tabcolsep}{5pt} 
    \scriptsize 
    \begin{tabular}{|c|c|c|c|c|c|c|c|c|}
        \hline
        \textbf{Slab} & \textbf{IoU} & \textbf{Precision} & \textbf{Recall} & \textbf{F1} & \textbf{FNR} & \textbf{FPR} & \textbf{TNR} & \textbf{ROC} \\
        \hline
        Slab1 & 0.87 & 0.94 & 0.91 & 0.93 & 0.08 & 0.38 & 0.61 & 0.77 \\
        Slab2 & 0.83 & 0.96 & 0.85 & 0.90 & 0.15 & 0.25 & 0.75 & 0.80 \\ 
        Slab3 & 0.80 & 0.96 & 0.83 & 0.89 & 0.17 & 0.25 & 0.75 & 0.79 \\ 
        Slab4 & 0.82 & 0.96 & 0.85 & 0.90 & 0.15 & 0.26 & 0.74 & 0.79 \\ 
        Slab5 & 0.89 & 0.94 & 0.93 & 0.94 & 0.07 & 0.43 & 0.57 & 0.75 \\ 
        Slab6 & 0.87 & 0.93 & 0.94 & 0.93 & 0.06 & 0.57 & 0.43 & 0.68 \\ 
        Slab7 & 0.89 & 0.95 & 0.93 & 0.94 & 0.07 & 0.40 & 0.60 & 0.77 \\ 
        Slab8 & 0.90 & 0.94 & 0.95 & 0.95 & 0.05 & 0.43 & 0.57 & 0.76\\
        \hline

    \end{tabular}   
    \caption{Performance metrics for cluster maps}
    \label{table:Performance Metrics for Cluster Maps}
\end{table}

The performance metrics for both the binary contour maps and cluster maps highlight significant differences in the detection of defects across the slabs. Binary contour maps achieve high precision values (0.96 to 0.98), effectively minimizing false positives. However, the recall values, which range between 0.76 and 0.79, are comparatively lower, suggesting that some defect areas present in the ground truth were not identified. This is further reflected in the relatively higher FNR, which are consistently above 0.20 for all slabs. The AUC-ROC values for the binary contour maps range from 0.78 to 0.83, reflecting a moderate ability to differentiate between defect and non-defect regions.

In contrast, the cluster maps show improved overall performance. The IoU values, ranging from 0.82 to 0.90, indicate better overlap with the ground truth. Precision remains high (0.93 to 0.96), while recall values (0.83 to 0.94) are notably higher than those of the binary contour maps, resulting in lower FNR values (0.05 to 0.17). The higher recall demonstrates the ability of cluster maps to capture more defect areas. The F1 score, which balances precision and recall, is consistently higher for the cluster maps, indicating their robustness. However, the AUC-ROC values for the cluster maps remain similar, ranging from 0.68 to 0.80, showing only slight improvement in defect differentiation.

Overall, the cluster maps outperform the binary contour maps in terms of Recall, IoU, F1-score, and FNR, making them a more reliable approach for defect detection in concrete slabs. However, the moderate AUC-ROC values for both methods suggest room for improvement in accurately separating defect and non-defect regions.

\begin{figure}[t]
    \centering
    \includegraphics[width=\linewidth]{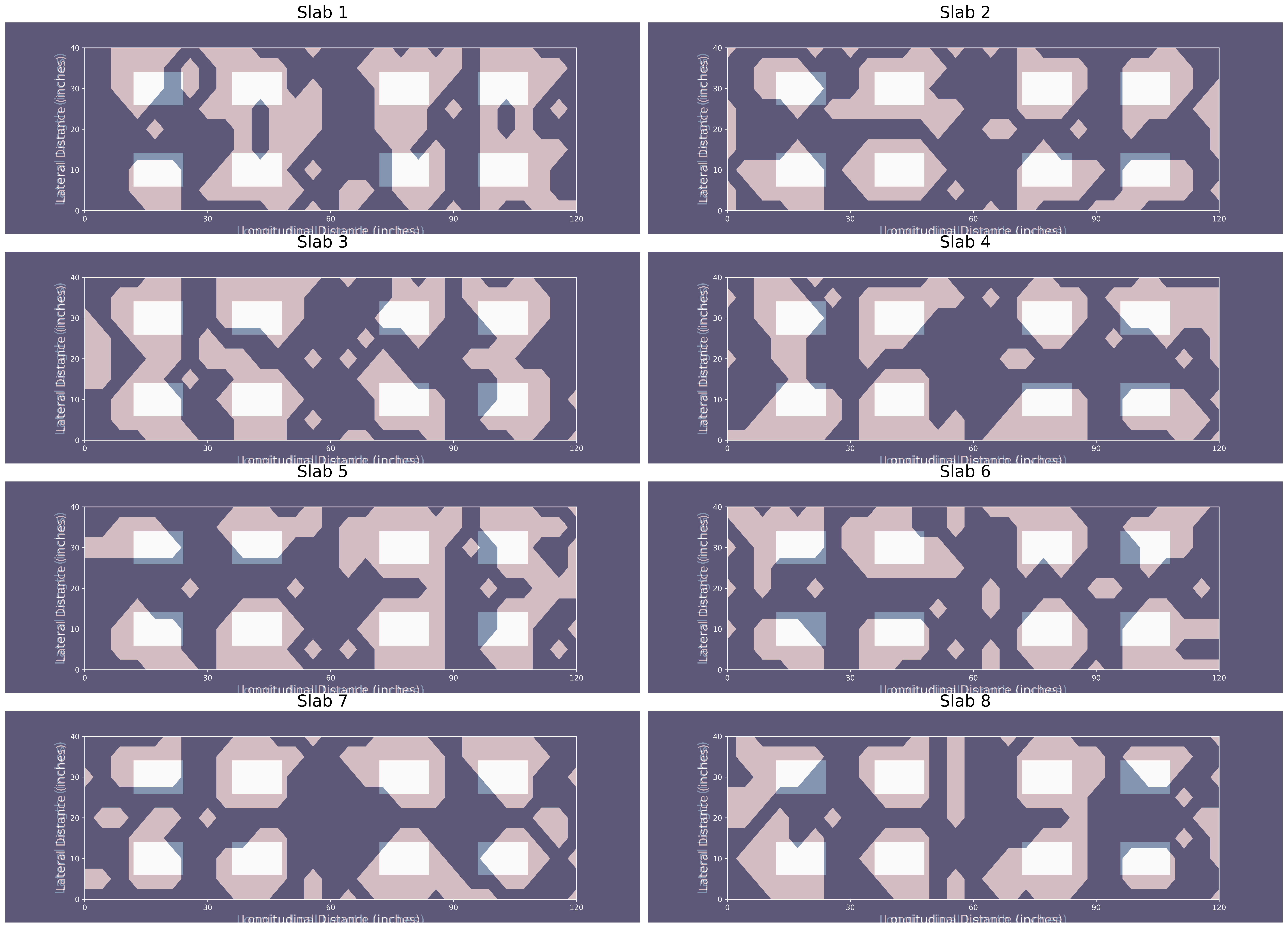}
    \caption{Overlay of top-view and binary maps}
    \label{figure:overlay_contour_image}
\end{figure}

\begin{figure}[t]
    \centering
    \includegraphics[width=\linewidth]{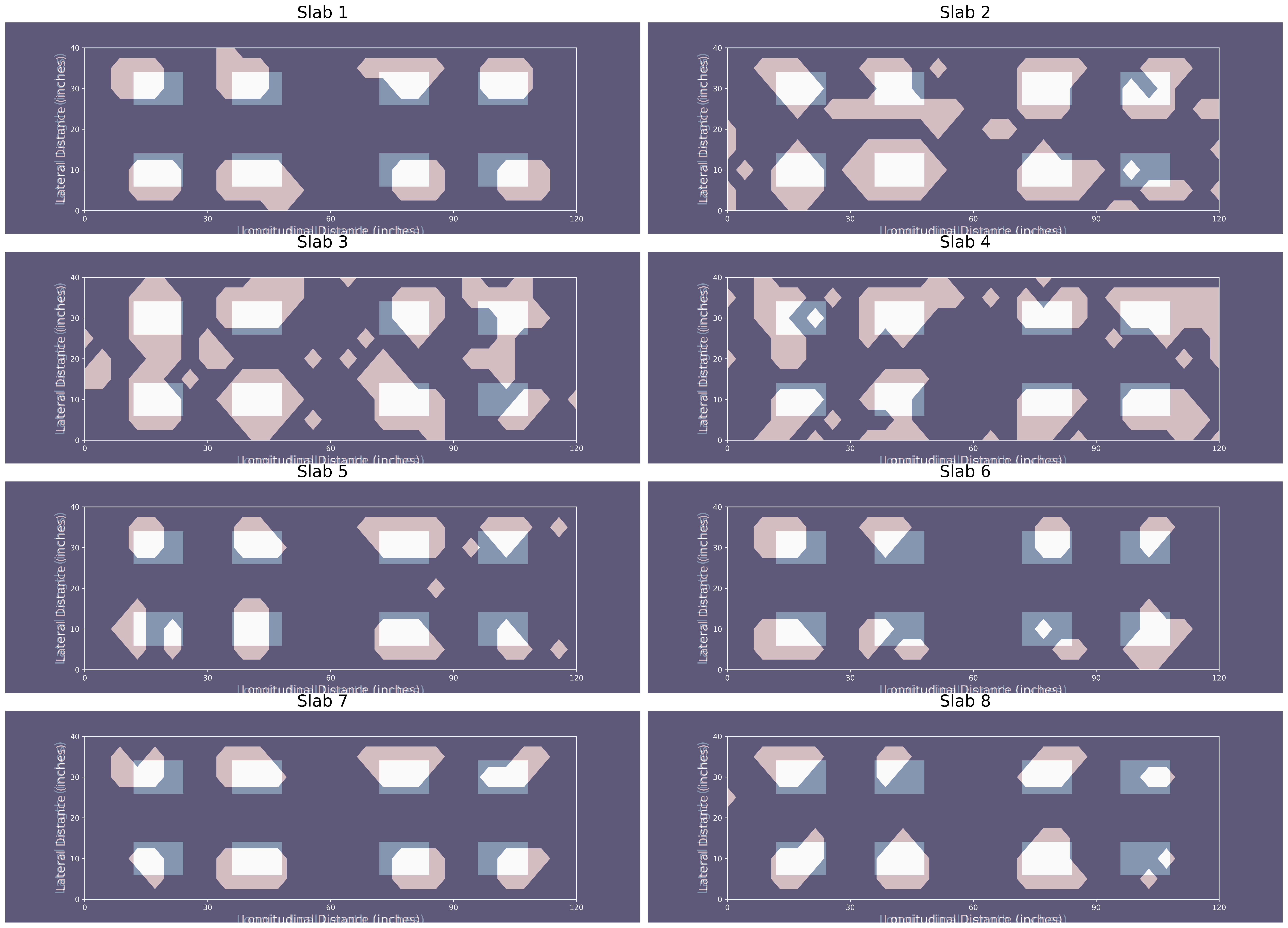}
    \caption{Overlay of top-view and cluster maps}
    \label{figure:overlay_cluster_image}
\end{figure}

The overlay images of the binary contour plot and cluster map for all slabs in Figure~\ref{figure:overlay_contour_image} and \ref{figure:overlay_cluster_image} effectively demonstrate the alignment between predicted defect areas and the ground truth (top view). In the binary contour plot, the defect regions captured are fairly consistent with the top view, showing an overlap but with slight mismatches around the edges of some regions. The cluster map, on the other hand, refines the defect predictions, providing better alignment with the top view while reducing noise and capturing clearer defect boundaries. The comparative precision and recall metrics support this visual observation, with the cluster map offering improved defect localization, as evidenced by its higher recall and IoU values compared to the binary contour plot. These visual and quantitative analyses together validate the effectiveness of the clustering approach for accurate defect detection in the concrete slab.

\subsection{Summary of Results}The evaluation metrics across all slabs indicate strong performance in defect detection for both binary contour plots and cluster maps. Key observations include:

\subsubsection{Binary Contour Plots} Binary contour plots deliver reliable defect detection with AUC-ROC values (0.78 to 0.83), Precision above 0.96, and F1-scores (0.84 to 0.87). Although IoU values (0.73 to 0.78) indicate moderate overlap with ground truth masks, the slightly higher false negative rates suggest that binary contour plots tend to miss more defects compared to cluster maps.

\subsubsection{Cluster Maps} Cluster maps achieve high precision (above 0.93), IoU values (0.80 to 0.90), and F1-scores (0.89 to 0.95), demonstrating superior defect localization and detection. Although their AUC-ROC values (0.68 to 0.80) are slightly lower than binary contour plots, they remain effective in identifying defect-prone areas.

Overall, cluster maps excel in IoU, recall, and F1-scores for better defect localization, while binary contour plots achieve higher AUC-ROC values, offering superior differentiation. Cluster maps effectively minimize missed defects, and binary contour plots reduce false positives.

\subsection{Error Analysis} 
Despite strong performance, some challenges remain: 

\subsubsection{Alignment Sensitivity} Resizing and alignment introduce distortions, which affect IoU and recall, particularly near edges or in subtle defect regions. 

\subsubsection{Subtle Defects} Low-frequency variations are occasionally missed, lowering recall; adaptive thresholds address this and can enhance performance.

\subsubsection{Threshold Sensitivity} Minor inaccuracies in dynamic thresholds impacted faint defect detection, requiring further refinement to improve sensitivity.

\subsection{Generalized Detection Patterns} The evaluation reveals consistent patterns in the identification of defect-prone areas: \subsubsection{Robust Defect-Prone Area Identification} High AUC-ROC, precision, and IoU values demonstrate the methodology's effectiveness in consistently identifying regions likely to contain defects, such as areas with delamination, voids, or other anomalies.

\subsubsection{Challenges with Subtle Variations} Detection accuracy slightly decreases for defect-prone areas with minimal frequency variations or complex spatial patterns, highlighting opportunities for improvement in frequency-adaptive analysis.

\subsubsection{Consistency Across Slabs} Similar evaluation metrics across slabs confirm the methodology's reliability and scalability for non-destructive evaluation of concrete slabs, particularly those without overlays.

Overall, cluster maps provide superior defect detection accuracy and localization compared to binary contour plots, making them the preferred approach for identifying defect-prone areas in concrete slabs.

\section{Conclusion}
The analysis of frequency distributions across the slabs highlights effectiveness of the IE method of identifying defect-prone and intact regions. This non-destructive, adaptive assessment enables precise structural health evaluations for targeted maintenance and repair.

The variability in defect-prone regions emphasizes the need for slab-specific, spatially resolved analysis. Adaptive thresholds and visualizations, such as contour and cluster maps, provide consistent and reliable detection, offering actionable insights for structural evaluation.

This study showcases the potential of frequency-based, adaptive methodologies for NDE of concrete slabs, facilitating targeted and cost-effective maintenance. The scalable approach focuses on identifying defect-prone areas and can be extended to slabs with overlays and other frequency-based signals, offering versatility for diverse NDE applications.

\section{Future Scope}
This study paves the way for advancements in concrete structure evaluation. Future research will refine defect detection algorithms using machine learning on larger datasets for improved automation and accuracy. Expanding the approach to other materials and structures, like bridges and foundations, will test its broader applicability. Combining additional NDE methods, such as Electric Resistivity and Ultrasonic Waves, with IE testing, enhances defect detection by offering complementary insights. Finally, developing real-time monitoring systems using sensors and frequency analysis will enable continuous structural integrity assessments, promoting safer and more resilient infrastructure

\section*{Acknowledgment}
The authors would like to thank the Federal Highway Administration (FHWA) for providing the dataset and detailed descriptions of the slabs used in this study. This work was made possible with their invaluable resources and support.

\end{document}